\documentclass{emulateapj}
\usepackage{color}
\usepackage{bm}
\usepackage{lineno}
\usepackage{amsmath}
\shorttitle{Electron--Positron Cosmic-Ray Propagation with the CALET Spectrum}
\shortauthors{Asano, et al.}

\begin{document}
\pagenumbering{arabic}
\title{
Monte Carlo Study of Electron and Positron Cosmic-Ray Propagation with the CALET Spectrum
}
\author{Katsuaki Asano\altaffilmark{1},
Yoichi Asaoka\altaffilmark{1}, Yosui Akaike\altaffilmark{2,3},
Norita Kawanaka\altaffilmark{4,5,6},
Kazunori Kohri\altaffilmark{7,8},
Holger M. Motz\altaffilmark{9},
Toshio Terasawa\altaffilmark{1}
}
\email{asanok@icrr.u-tokyo.ac.jp}

\affil{\altaffilmark{1}Institute for Cosmic Ray Research, The University of Tokyo,
5-1-5 Kashiwanoha, Kashiwa, Chiba 277-8582, Japan}
\affil{\altaffilmark{2}Waseda Research Institute for Science and Engineering, Waseda University, 17 Kikuicho, Shinjuku, Tokyo 162-0044, Japan}
\affil{\altaffilmark{3}JEM Utilization Center, Human Spaceflight Technology Directorate, Japan Aerospace Exploration Agency,
2-1-1 Sengen, Tsukuba, Ibaraki 305-8505, Japan}
\affil{\altaffilmark{4}Hakubi Center, Kyoto University, Yoshida Honmachi, Sakyo-ku, Kyoto 606-8501, Japan}
\affil{\altaffilmark{5}Department of Astronomy, Graduate School of Science, Kyoto University,
Kitashirakawa Oiwake-cho, Sakyo-ku, Kyoto 606-8502, Japan}
\affil{\altaffilmark{6}Yukawa Institute for Theoretical Physics, Kyoto University, Kitashirakawa Oiwake-cho, Sakyo-ku, Kyoto 606-8502, Japan}
\affil{\altaffilmark{7}Theory Center, IPNS, KEK, and SOKENDAI, 1-1 Oho, Tsukuba, Ibaraki 305-0801, Japan}
\affil{\altaffilmark{8}Kavli IPMU (WPI), UTIAS, The University of Tokyo, Kashiwa, Chiba 277-8583, Japan}
\affil{\altaffilmark{9}Faculty of Science and Engineering, Global Center for Science and Engineering,
Waseda University, 3-4-1 Okubo, Shinjuku, Tokyo 169-8555, Japan}

\begin{abstract}
Focusing on the electron and positron spectrum measured with CALET,
which shows characteristic structures, we calculate flux contributions
of cosmic rays escaped from supernova remnants, which were randomly born.
We adopt a Monte Carlo method to take into account the stochastic property
of births of nearby sources. We find that without a complicated energy dependence
of the diffusion coefficient, simple power-law diffusion coefficients
can produce spectra similar to the CALET spectrum even with a dispersion in the injection index.
The positron component measured with AMS-02 is consistent with a bump-like structure
around 300 GeV in the CALET spectrum.
One to three nearby supernovae can contribute up to a few tens of percent
of the CALET flux at 2--4 TeV, ten or more
unknown and distant ($\gtrsim 500$ pc) supernovae
account for the remaining several tens of percent of the flux.
The CALET spectrum, showing a sharp drop at $\sim 1$ TeV,
allows for a contribution of cosmic rays from an extraordinary event
which occured $\sim 400$ kyr ago.
This type of event releases electrons/positrons with a total energy more than 10 times
the average energy for usual supernovae,
and its occurrence rate is lower than $1/300$ of the usual supernova rate.
\end{abstract}


\maketitle

\section{Introduction}
\label{sec:intro}

Direct measurements of electron and positron cosmic-rays (CRs)
have been pioneered by various teams such as BETS \citep{BETS},
HEAT \citep{HEAT}, ATIC \citep{ATIC}, PPB-BETS \citep{PPB},
and PAMELA \citep{PAMELA}.
Recently, the electron and positron CR study
(total CR flux not distinguishing their charge signs)
has been greatly advanced with new instruments
such as AMS-02 \citep{ams14,ams19},
{\it Fermi}-LAT \citep{abd17}, CALET \citep{cal17,cal18,epmCALET},
and DAMPE \citep{dampe}.
While the spectra of electron and positron CRs measured with
AMS-02 and CALET are consistent,
the spectra obtained with {\it Fermi}-LAT and DAMPE
show slightly harder shape and higher fluxes above 300 GeV.
The discrepancy between the spectra reported by the different teams
has not been resolved yet.

The CALET and DAMPE collaborations reported the spectra as far as 4.6--4.8 TeV
\citep{dampe,cal18}.
The DAMPE team claimed the detection of
a spectral break at $\sim$TeV \citep{dampe},
and the CALET spectrum also suggests a softening above $\sim$ TeV \citep{cal18}.
The recent preliminary results with CALET
also confirm a broken power-law at $\sim$TeV with higher statistics \citep{epmCALET}.
The TeV break can be attributed to the radiative cooling effect \citep{lip19,mer18},
or to electron--positron pairs coming from a single nearby pulsar wind nebula (PWN)
like Monogem \citep{din21}.
\citet{kis12} and \citet{shi19} proposed models with millisecond pulsars
as sources of CR components responsible for the TeV break.

As the cooling timescale for TeV electrons is short,
the sources of TeV electron/positron CRs must be nearby.
Possible sources are known pulsars like the Vela pulsar \citep{hua18},
unknown multiple pulsars \citep{kaw11,cho18}, 
known supernova remnants (SNRs) like Vela \citep{kob04}
or Cygnus Loop \citep{man19},
unknown old SNRs \citep{fan18,for20},
and white dwarf pulsars \citep{kas11}.
The small number of nearby sources implies a large dispersion
in the predicted TeV flux as discussed in \citet{kaw10,cho18,fan18,mer18}.

As a conservative approach, the difference between DAMPE and CALET spectra
could be treated as systematic errors.
In this paper, however, we focus on only the CALET spectrum with its
characteristic structures: a spectral bump around 200--300 GeV,
and a sharp flux drop at $\sim 1$ TeV.
Since the CALET and AMS-02 spectra of electron and positron CRs are consistent,
it may be reasonable to discuss the CALET electron and positron spectrum
together with the positron spectrum measured with AMS-02 \citep{ams14pos,ams19}.
The positron component above 100 GeV is apparently a different component
from the electron component, which dominates the electron and positron flux.
Potential sources of positrons above 100 GeV are pulsars \citep{cho18,di21,oru21}
or a single SNR interacting dense medium \citep{fuj09,koh16}.
The spectral bump in the CALET spectrum may be attributed to the
positron component.

The dominant component of electron and positron CRs
are electrons from SNRs.
To investigate the effect of the number fluctuation of nearby sources,
we adopt a Monte Carlo method in our calculations
\citep[see also][]{mer18,evo21,oru21}.
We examine whether SNRs, which randomly occur, can reproduce
the characteristic structures in the CALET spectrum,
or if different origins are required.
Compared to the previous study by \citet{evo21},
we focus more on the modulation of the CALET spectrum.
We also study the influence of the variation in the injection spectra
on the total spectrum.
The dispersion in the injection index was taken into account
for pulsars in \citet{cho18}.
In our study, the variation of the injection index
is introduced for electron CRs from SNRs as well.
A very large dispersion in the injection spectral index
may contradict the observed spectrum.

In this paper, we adopt the officially published data
of the electron and positron spectrum with CALET \citep{cal18},
which is consistent with the preliminary updated result of
\citet{epmCALET}.

\section{Numerical Methods}
\label{sec:method}

We calculate the CR flux at the earth by integrating the contributions
of individual sources.
We use the standard Green's function method to
estimate the contribution of a single source \citep{ato95}.
First, we review the calculation method for a single source.

\subsection{CR Propagation}

The number density spectrum of electrons/positrons $n(\gamma)$
from a single source at the origin $r=0$ in spherical coordinates
is calculated with the diffusion equation,
\begin{equation}
\frac{\partial n}{\partial t}-K \frac{1}{r^2}\frac{\partial}{\partial r}
\left( r^2 \frac{\partial n}{\partial r} \right)
+ \frac{\partial}{\partial \gamma}
\left( \dot{\gamma} n \right)=Q_{\rm inj}(\gamma),
\label{basiceq}
\end{equation}
where $Q_{\rm inj}$ is the injection term,
and the diffusion coefficient in the second term is assumed as
a power-law function of energy,
\begin{equation}
K(\gamma)=K_0 \left( \frac{\gamma m_{\rm e}c^2}{\mbox{GeV}} \right)^q.
\end{equation}
In the simplest model with isotropic Alfv\'en
turbulence, the index $q$ is related to the power-law index for
the power spectrum of the turbulence in the interstellar medium.
The standard Kolmogorov turbulence ($q=1/3$)
and Iroshnikov--Kraichnan turbulence ($q=1/2$) are representative models.

The radiative cooling effect in the third term of equation (\ref{basiceq})
includes synchrotron (SYN) cooling and inverse Compton (IC) cooling as
\begin{equation}
\dot{\gamma} m_{\rm e}c^2=-b_{\rm IC}(\gamma)-b_{\rm SYN}(\gamma).
\end{equation}
The synchrotron cooling term is written as
\begin{eqnarray}
b_{\rm SYN}(\gamma)=\frac{4}{3}\sigma_{\rm T} c \frac{B^2}{8 \pi} \gamma^2,
\end{eqnarray}
where $\sigma_{\rm T}$ is the cross section of the Thomson scattering.
Below we adopt the magnetic field of $B=3\mu$G.
For the IC cooling term $b_{\rm IC}$, we take into account the Klein--Nishina effect
as explained below.

When an electron with Lorentz factor $\gamma$ is in a photon field
of the spectral density $n_{\rm ph}(\varepsilon_0)$,
the photon production rate of an energy $\varepsilon_1$
via IC is
\begin{eqnarray}
\dot{N}(\varepsilon_0,\varepsilon_1)&&=\frac{3 \sigma_{\rm T}c}{4 \varepsilon_0 \gamma^2}
n_{\rm ph}(\varepsilon_0) \nonumber \\
&\times& \left[ 2q \ln q+(1+2q)(1-q)+\frac{(\Gamma q)^2}{2(1+\Gamma q)}(1-q) \right],
\nonumber \\
\end{eqnarray}
where
\begin{eqnarray}
q \equiv \frac{x_1}{\Gamma (1-x_1)}, \quad x_1 \equiv \frac{\varepsilon_1}{\gamma m_{\rm e} c^2},
\quad \Gamma \equiv \frac{4 \gamma \varepsilon_0}{m_{\rm e} c^2},
\end{eqnarray}
\citep[e.g.][]{blu70,del10}.
With this photon production rate, the cooling rate via IC is written as
\begin{eqnarray}
b_{\rm IC}(\gamma)=\int d \varepsilon_0 \int d \varepsilon_1
(\varepsilon_1-\varepsilon_0) \dot{N}(\varepsilon_0,\varepsilon_1).
\end{eqnarray}
This is rewritten with
$x_1=\Gamma q/(1+\Gamma q)$ and $dx_1=\Gamma/(1+\Gamma q)^2 dq$ as
\begin{eqnarray}
b_{\rm IC}(\gamma)&=& \int d \varepsilon_0 \int_{1/(4 \gamma^2)}^1 d q
\frac{\Gamma^2 (\gamma m_{\rm e} c^2)^2}{(1+\Gamma q)^2} \nonumber \\
&\times&
\left( \frac{q}{1+\Gamma q}-\frac{1}{4 \gamma^2} \right)
\dot{N}(\varepsilon_0,\varepsilon_1).
\end{eqnarray}

We consider a case in which
the photon field is approximated by the diluted Planck distribution,
\begin{eqnarray}
n_{\rm ph}(\varepsilon_0)=\frac{15 U_{\rm ph}}{(\pi T)^4} \frac{\varepsilon_0^2}
{\exp{(\varepsilon_0/T)}-1},
\end{eqnarray}
where $U_{\rm ph}$ and $T$ are the photon energy density and temperature, respectively.
In this case, the Klein--Nishina effect is separately
represented by a dimensionless factor $f_{\rm KN}$ as
\begin{eqnarray}
b_{\rm IC}(\gamma)=\frac{4}{3}\sigma_{\rm T} c U_{\rm ph} \gamma^2
f_{\rm KN}(\gamma,T).
\end{eqnarray}
Defining dimensionless quantities
$y=\varepsilon_0/T$ and $z=4 \gamma T/(m_{\rm e} c^2)$,
we can write $\Gamma=yz$,
then the factor $f_{\rm KN}$ is expressed as a function of $z$ as
\begin{eqnarray}
f_{\rm KN}(z)&=& \frac{135}{\pi^4}
\int d y \frac{y^3} {\exp{(y)}-1} \nonumber \\
&&\times \int_{1/(4 \gamma^2)}^1 d q 
\frac{1}{(1+yz q)^2} \left( \frac{q}{1+yz q}-\frac{1}{4 \gamma^2} \right) \nonumber \\
&&\times \left[ 2q \ln q+(1+2q)(1-q)+\frac{(yz q)^2}{2(1+yz q)}(1-q) \right]. \nonumber \\
\end{eqnarray}
While $f_{\rm KN}(z)\simeq 1$ for $z \ll 1$,
\begin{eqnarray}
f_{\rm KN}(z) \simeq \frac{45}{4 \pi^2 z^2}
\left( \ln{z}-1.9805 \right),
\end{eqnarray}
for $z \gg 1$.
For the numerical approximation of the function $f_{\rm KN}(z)$,
we use the fitting formulae in \citet{fan21} in our numerical calculation.

Following \citet{del10,evo20},
the photon field in the interstellar space
is modeled by a sum of 6 diluted Planck distributions:
$T=2.725,~33.07,~313.32,~3249.3,~6150.4$, and $23209.0$ K
with $U_{\rm ph}=0.26,~0.25,~0.055,~0.37,~0.23$, and $0.12~\mbox{eV}~\mbox{cm}^{-3}$,
respectively.

\subsection{CR Injection}
\label{CRinj}

Let us assume the power-law plus exponential cutoff
for the injection spectrum from a CR source at a distance $r_i$
and the CR release time $t=-t_i$ as
\begin{equation}
Q_{\rm inj}(\gamma)=Q_i \left( \frac{\gamma m_{\rm e}c^2}{\mbox{GeV}} \right)^{-\alpha}
\exp{\left( -\frac{\gamma}{\gamma_{\rm max}}\right)}
\delta(\bm{r}-\bm{r}_i)\delta(t+t_i),
\label{eq:inj}
\end{equation}
which is expressed by three parameters,
the normalization $Q_i$, the spectral index $\alpha$,
and the cutoff Lorentz factor $\gamma_{\rm max}$.

Here, we approximate the injection process by prompt injections.
Note that the spectral shape in equation (\ref{eq:inj})
is an effective one after the escape process from the source.
The spectral shape of escaped CRs largely depends on the escape and the cooling processes,
which are observationally and theoretically uncertain.
Although the injection is generally time-dependent,
we approximate the injection process using equation (\ref{eq:inj})
as a CR spectrum at the escape time $t=-t_i$.

At the earth, multiple sources with different distances $r_i$
and ages $t_i$ contribute to the electron and positron CR spectrum.
To integrate these contributions, we carry out Monte Carlo simulations
for a distribution of sources in distance and age.
The primary sources of electron cosmic-rays are SNRs.
The total energy of CR electrons escaping from a SNR,
\begin{equation}
E_{\rm tot}=\int_{\gamma_{\rm m}}^\infty d \gamma Q_{\rm inj}(\gamma) \gamma m_{\rm e}c^2,
\end{equation}
is fixed to $9.0 \times 10^{47}$ erg,
which is about $0.1$ \% of the explosion energy of a supernova (SN),
and $1$ \% of the CR proton energy.
The total CR energy is defined by integrating above 1 GeV,
so that the lower bound in the above integral is $\gamma_{\rm m} \simeq 1960$.

We also consider pulsars as sources of electron--positron CRs.
Several models of a significant contribution
from Geminga pulsar have been proposed as explanation for the spectral structure at $\sim$ TeV.
The present spin-down luminosity $L_{\rm SD}=3.5 \times 10^{34}~\mbox{erg}~\mbox{s}^{-1}$
and the characteristic age $t_{\rm age}=3.0 \times 10^5$ yr of the Geminga pulsar \citep{ber92}
imply an energy release of $L_{\rm SD} t_{\rm age}=3.3 \times 10^{47}$ erg.
The required energy release in the Geminga models \citep[e.g.][]{fan18}
is $\sim 10^{49}$ erg, which may require a prompt energy release just at the pulsar's
birth.

If electron--positron pairs freely escape from the Geminga pulsar,
it can account for the AMS-02 positron component \citep{yus09}.
However, such an efficient escape from PWNe
is rejected for Vela X \citep{hua18}; the CR flux from Vela X
would largely exceed the observed flux in this case.
In addition, many pulsars were born after Geminga's birth,
and they should similarly contribute to the positron CR flux.
If we adopt a similar model parameter set for other pulsars,
the total positron flux may exceed the observed flux.

Gamma-ray observations show that the diffusion coefficient around the Geminga pulsar
is largely suppressed \citep{abe17}.
In general, PWNe are surrounded by SNRs,
and the CR energy density around nebulae is much higher than
the average value in the interstellar space,
which may be reasons for the suppressed diffusion coefficient.
A large fraction of CRs released from Geminga may remain around the source yet.
If we simply adopt the suppressed diffusion coefficient in \citet{abe17}
($K_{\rm G}=3.2\times 10^{26}~\mbox{cm}^2~\mbox{s}^{-1}$ at 100 GeV)
and the gamma-ray halo size ($R_{\rm G}\sim 50$ pc),
the escape time of 100 GeV electrons from the Geminga halo
is $R_{\rm G}^2/(4 K_{\rm G}) \simeq 590$ kyr, which is longer than
the age of Geminga.
Although this estimate may be too pessimistic,
we can assume a significant time-delay for the CR escape from PWNe.

Secondary CRs produced from proton CRs propagating in the interstellar medium
are also the source of electron and positron CRs.
Compared to the electron and positron flux,
the contribution of such secondary CRs is negligible above 10 GeV \citep[see e.g.,][]{di21}.
Even for the positron spectrum, the secondary CRs can be neglected
above $\sim 50$ GeV,
so that we omit the secondary electrons and positrons in this paper.

\subsection{CR Flux at the Earth}

Since the elapsed time to evolve from $\gamma_0$ to $\gamma$ via radiative cooling
is numerically calculated by
\begin{eqnarray}
t_i(\gamma,\gamma_0)=\int_\gamma^{\gamma_0} d \gamma' \frac{m_{\rm e}c^2}{b(\gamma')},
\end{eqnarray}
we can produce a table of the initial Lorentz factor $\gamma_0(\gamma,t_i)$
as a function of the present Lorentz factor $\gamma$
and the elapsed time $t_i$.
We also produce a table for the diffusion radius defined as
\begin{eqnarray}
r_{\rm dif}(\gamma,t_i)=2 \sqrt{\int_\gamma^{\gamma_0}
d \gamma' \frac{m_{\rm e}c^2 K(\gamma')}{b(\gamma')}}.
\end{eqnarray}
Using those tables numerically prepared in advance, we obtain the CR spectrum at the earth
from a single source \citep{ato95} as
\begin{eqnarray}
n_i(\gamma)&=&\frac{Q_i}{\pi^{3/2} r_{\rm dif}^3}\frac{b(\gamma_0)}{b(\gamma)}
\left( \frac{\gamma_0 m_{\rm e} c^2}{\mbox{GeV}} \right)^{-\alpha} \nonumber \\
&& \times \exp{\left( -\frac{\gamma_0}{\gamma_{\rm max}}\right)}
\exp{\left( -\frac{r_i^2}{r_{\rm dif}^2}\right)},
\end{eqnarray}
\citep[see also e.g.,][]{hoo17}.

\begin{figure}[!h]
\centering
\epsscale{1.1}
\plotone{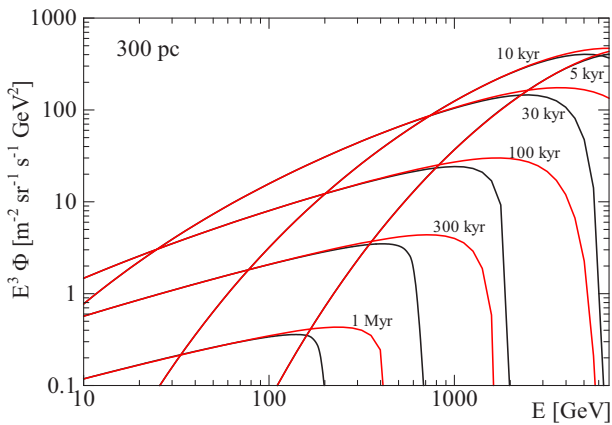}
\caption{Electron spectra of a single source
at $r=300$ pc that released electrons 5 kyr,
10 kyr, 30 kyr, 100 kyr, 300 kyr, and 1 Myr ago.
The red and black lines show the results with and
without the Klein--Nishina effect, respectively.
The parameters are $K_0=3.0 \times 10^{28}~\mbox{cm}^2~\mbox{s}^{-1}$,
$q=1/3$, $E_{\rm tot}=10^{48}$ erg,
$\alpha=2.0$, and $\gamma_{\rm max} m_{\rm e} c^2=10$ TeV.
\label{fig:test}}
\end{figure}

As we explained above, the Klein--Nihina effect
is numerically included in our method.
Figure \ref{fig:test} shows the results of test calculations
to comprehend the Klein--Nishina effect.
If we neglect the Klein--Nishina effect,
the cutoff energy is underestimated and the cutoff shape
becomes sharper than the exact spectrum.
Around a few TeV, the contributions from sources
occurred $\sim 100$ kyr before are drastically different
between the calculations with and without the Klein--Nishina effect.

Though \citet{evo21} take into account the spiral structure of our galaxy
for the source distribution, we simply assume constant and homogeneous rates
of SNe in this paper.
The scale height $h$ is set as $\pm 300$ pc, in which the SN rate
is constant, and we neglect the contributions above or below this scale height.
We integrate sources with the radial distance $R \leq 10$ kpc
from the earth and $t_i \leq 10^8$ yr.
The final electron and positron spectrum at the earth is calculated
by summing up all the contributions of SNe as
\begin{equation}
\Phi(E)=\frac{c}{4 \pi}\sum_i \frac{n_i(\gamma)}{m_{\rm e}c^2},
\end{equation}
in unit of $\mbox{m}^{-2} \mbox{sr}^{-1} \mbox{s}^{-1} \mbox{GeV}^{-1}$.

\section{Monte Carlo Results}
\label{MCres}

First, we assume the Kolmogorov-type diffusion ($q=1/3$),
which is consistent with the B/C ratio observation
with AMS-02 \citep{ams16,ams18} and the preliminary
CALET result \citep{B/CCALET}, though values
$q=0.41$--0.44 \citep{gen15,kap15} or a broken power-law model
\citep{evo19} are
acceptable for the combined fit to the B/C ratio and heavy nuclei.
The normalization is assumed as
$K_0=3.0 \times 10^{28}~\mbox{cm}^2~\mbox{s}^{-1}$ \citep{str07}.

As will be shown below, the Monte Carlo results show large variations
of the spectral shape. For a certain parameter set, even when most of the results
do not agree with the observed data, we may find a few percent of results
consistent with the data.
Considering this large dispersion in the Monte Carlo results
and the simplified assumptions in our theoretical model,
we do not quantitatively estimate how models agree with the data,
as it would hardly yield useful information.

In \S \ref{sub1}, we test the simplest model; only one type of sources
with a common injection index is considered.
We obtain typical parameter values agreeing with the
CALET spectrum.
However, the spectral bump at 100--300 GeV and the sharp drop at $\sim 1$ TeV
in the CALET spectrum require other types of sources
with a different energy release and injection index.
In \S \ref{sub2},
we add two components, one of which is consistent with the positron component
measured with AMS-02, and another one for reproducing the sharp drop at $\sim 1$ TeV
originating from an extraordinary source with a hard injection index.
In \S \ref{sub3}, we test cases with a dispersion in the injection index
for SNR sources. Our results roughly provide an allowed width of
the injection index distribution.
Based on this model, we also discuss the number of nearby sources
contributing to the measured flux above 3 TeV.
In \S \ref{IK}, for reference, we show results for
the Iroshnikov-Kraichnan type diffusion.
The difference in the diffusion index does not change the main conclusions.

\subsection{Constant Index}
\label{sub1}

\begin{figure}[!h]
\centering
\epsscale{1.1}
\plotone{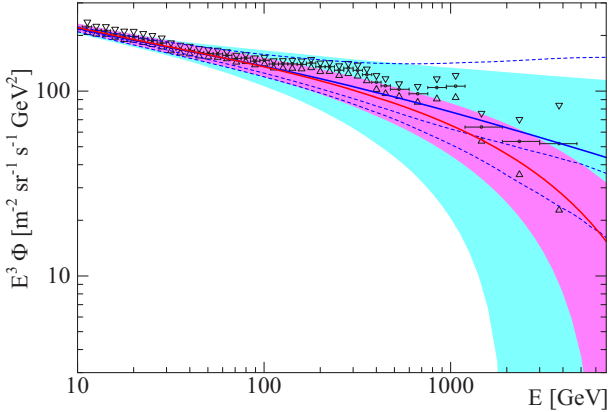}
\caption{The Monte Carlo simulation results
with a constant injection index $\alpha=2.58$.
The other parameters are
SN rate of $7.7 \times 10^{-2}~\mbox{kpc}^{-2}~\mbox{kyr}^{-1}$,
$K_0=3.0 \times 10^{28}~\mbox{cm}^2~\mbox{s}^{-1}$,
$q=1/3$, and $E_{\rm tot}=9.0 \times 10^{47}$ erg.
The maximum energy is assumed as infinity (blue)
and $\gamma_{\rm max} m_{\rm e} c^2=10$ TeV (red).
The observed data with CALET is taken from \citet{cal18}..
The thick lines are the average spectrum of 1000 trials.
The blue dashed lines are three examples of the trials
for $\gamma_{\rm max}=\infty$.
The spectral variation ($1\sigma$) is shown
with cyan and magenta regions, for blue and red average spectra,
respectively.
\label{fig:No-disp}}
\end{figure}

In this section, we show the results
with a constant index of $\alpha$ as the simplest case.
In Figure \ref{fig:No-disp}, we summarize the Monte Carlo results
of 1000 trials with $\alpha=2.58$.
Though the injection parameters are fixed, the diversity
of the spatial and age distributions leads to
a variation in the spectral shape as shown by the blue dashed lines.
A significant fraction of the trials and the average spectrum (the blue solid line)
are consistent with the CALET electron and positron spectrum.
When accidentally a few recent and nearby sources occur in a sample,
the flux above $\sim$ TeV becomes significantly larger than
the CALET flux.
Such an example is the one with the highest flux among
the three dashed lines in Figure \ref{fig:No-disp}.
The variance of the spectrum is shown with the shaded region in the figure.

The assumed SN rate of
${\cal R}=7.7 \times 10^{-2}~\mbox{kpc}^{-2}~\mbox{kyr}^{-1}$
corresponds to 1 SN per 510 yr within 3 kpc
or equivalently 1 SN per 46 yr within 10 kpc
for our choice of $E_{\rm tot}=9.0 \times 10^{47}$ erg.
As the SN rate and $E_{\rm tot}$ are degenerate parameters,
the essential parameter is the energy injection rate density of CR electrons,
which is ${\cal R} E_{\rm tot}=6.9 \times 10^{46}~\mbox{erg}~\mbox{kpc}^{-2}~\mbox{kyr}^{-1}$.
A larger ${\cal R}$ holding the constant value of ${\cal R} E_{\rm tot}$
reduces the dispersion of the spectrum.
However, a much larger value of ${\cal R}$ is not realistic.

The spectral index $2.81 \pm 0.03$ of low-energy ($<300$ GeV) protons
measured with CALET \citep{cal19} and our injection index $\alpha=2.58$
is roughly consistent with the similar injection indices for protons and electrons
($2.81 \pm 0.03-q=2.48 \pm 0.03$ for protons) in this Kolmogorov case.

Figure \ref{fig:No-disp} shows two cases with and without the high-energy cutoff
in the injection spectrum.
The average spectra based on 1000 trials seem to prefer
the model with the cutoff at 10 TeV.
The model with the cutoff results in a narrower dispersion
in the spectrum variance (compare the cyan and magenta regions in the figure).
If a very nearby source was recently born, it dominates the CR flux above $\sim$ TeV.
Depending on the distance and age of such sources, the nearby sources
produce a large dispersion in the spectrum.
However, the cutoff in the spectra effectively reduces the relative
contribution of the nearby sources, which reduces the dispersion as shown in the figure.
As shown by the blue dashed lines,
we can find several examples consistent with the observed data
in our Monte Carlo trials without the cutoff.
We cannot reject the model without cutoff at a level of at least 1 $\sigma$.

\subsection{Contribution of Electron--Positron Pair Sources}
\label{sub2}

Though the single component model in \S \ref{sub1} roughly
reproduces the CALET spectrum, we do not find a result with a spectral modulation
similar to the characteristic structures
in the CALET spectrum, the 300 GeV bump and the $\sim 1$ TeV sharp drop,
in our 1000 trials.
Although those structures can be due to the statistical fluctuation,
we try to reconcile them by adding other components in this section.

The bump-like structure around 300 GeV can be related to the positron
component measured with AMS-02 \citep{ams19}.
One of the most promising candidates for the positron sources
are pulsars \citep[e.g.][]{zha01,gri07,hoo09,mal09,del10,cho18,di21,oru21}.
On the other hand, the sudden drop seen at $\sim 1$ TeV in the CALET spectrum
may be attributed to
a single source with a hard injection spectrum.
The sharp structure may correspond to the spectral cutoff due to radiative cooling
\citep{fuj09,kaw10,cho14,koh16,rec19}.
In this scenario, the cutoff energy at $\sim 1$ TeV implies a prompt CR injection
$\sim400$ kyr ago.

As will be shown below, the single source required for the TeV sharp drop
needs to release $\sim 10^{49}$ erg as electrons and/or positrons.
A very special source such as a hypernova remnant may directly accelerate
electrons of $\sim 10^{49}$ erg, which is much larger than the $E_{\rm tot}$
we have adopted for the usual SNR CRs.
Alternatively, we can consider an electron--positron pair source
such as a SNR interacting with molecular clouds
\citep{fuj09,koh16}, or pulsars.
The extrapolation from the positron spectrum with AMS-02
is not consistent with the TeV sharp drop.
Models with a single electron--positron pair source predict
that another component appears around 1 TeV in the positron spectrum.
The extra component contributes to the AMS-02 positron flux at 800 GeV,
unless the spectrum is extremely hard.
We cannot uniquely divide the measured positron spectrum
into the background due to pulsars and a nearby single component.

\begin{figure}[!h]
\centering
\epsscale{1.1}
\plotone{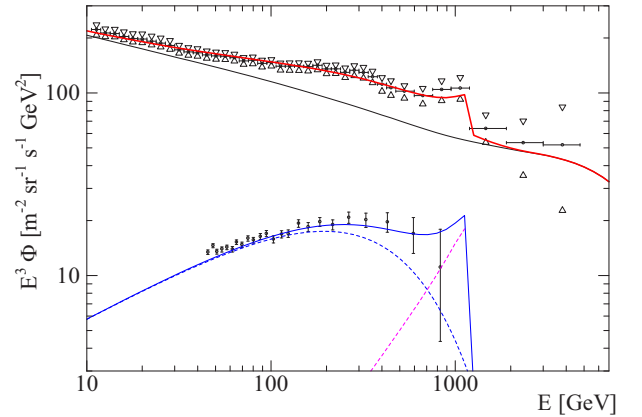}
\caption{An example of model spectra with
additional components consistent with
the positron spectrum measured by
AMS-02 \citep{ams19}.
The red line is the sum of the electron--positron pair component
(double the positron flux denoted with a blue solid line)
and the SN electron component (black solid),
which is a realization of the same Monte Carlo simulations
as in Fig. \ref{fig:No-disp} with $\gamma_{\rm max} m_{\rm e} c^2=10$ TeV.
The positron component (blue solid) is the sum of
two components denoted with dashed lines.
The blue dashed line is a positron component
of $\Phi \propto E^{-2.45} \exp{(-E/(350\mbox{GeV}))}$,
and the magenta dashed line is calculated
from an electron--positron pair ejection which occurred $420$ kyr ago at $r=1$ kpc
(source ``S''),
whose parameters are $E_{\rm tot}=1.5 \times 10^{49}$ erg,
$\alpha=1.5$, and $\gamma_{\rm max} m_{\rm e} c^2=100$ TeV.
\label{fig:synth}}
\end{figure}

Figure \ref{fig:synth} shows a synthesized spectrum
of three components: one example of the electron spectrum
chosen from the trials in the previous section (black solid),
the pair component from pulsars (blue dashed for only positrons),
and a hard pair component from a single nearby source
(source ``S'', magenta dashed for only positrons).
The total positron spectrum of the two latter pair components
is consistent with the AMS-02 positron spectrum,
and the total electron plus positron spectrum agrees with the CALET spectrum,
including the bump around 200--300 GeV
and the sharp drop at $\sim 1$ TeV.
In the first 100 trials in our Monte Carlo simulations
for electron CRs from SNe, we can find a few similar examples
to that in Figure \ref{fig:synth}.
Therefore, this example is not an extremely accidental case.

\begin{figure}[!h]
\centering
\epsscale{1.1}
\plotone{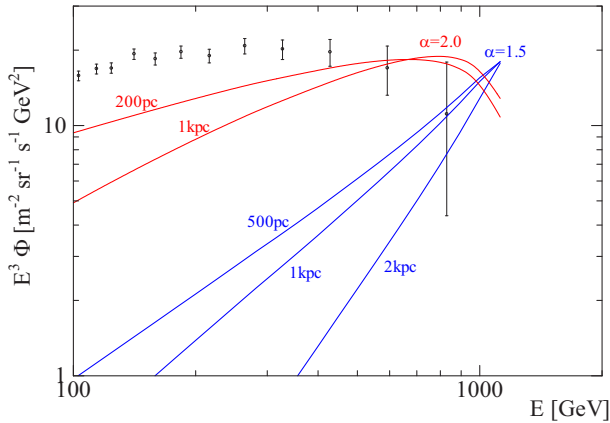}
\caption{Model spectra of positrons from a single source
$420$ kyr ago (source ``S'')
with $K_0=3.0 \times 10^{28}~\mbox{cm}^2~\mbox{s}^{-1}$,
$q=1/3$, and $\gamma_{\rm max} m_{\rm e} c^2=100$ TeV.
The blue lines are models of $\alpha=1.5$
from various distances as denoted,
for which the energy range is
$E_{\rm tot}=1.1$--$5.1 \times 10^{49}$ erg.
The red lines are models of $\alpha=2.0$,
for which the energy range is
$E_{\rm tot}=1.7$--$3.0 \times 10^{49}$ erg.
\label{fig:single-pos}}
\end{figure}

To reproduce the sharp drop at $\sim 1$ TeV, in Figure \ref{fig:synth},
we inject pairs from the source ``S'' with an index of $\alpha=1.5$
with the background pulsar component, for which we artificially assume
a spectral shape of $\Phi \propto E^{-2.45} \exp{(-E/(350\mbox{GeV}))}$.
If we change the source property of the source ``S'',
the background pulsar component should have another spectral shape.
As shown in Figure \ref{fig:single-pos},
a softer injection spectrum of $\alpha=2.0$
makes a spectral bump just below the cutoff energy at TeV,
which is partially due to the Klein--Nishina effect.
Softer injection spectra with $\alpha=2.0$ for the single source
are hard to reconcile with the sharp drop in the CALET spectrum,
and the contribution to the AMS-02 positron spectrum
seems too large as shown in Figure \ref{fig:single-pos}.
The required total energy is $1.1$--$5.1 \times 10^{49}$ erg
for a distance of $0.5$--$2$ kpc.

If the single source produced only electron CRs (e.g. electron acceleration
at a hypernova remnant), the positron component can be attributed to only usual pulsars.
In this case, we do not need to assume a different spectral shape
from the AMS-02 spectrum as we did in Figure \ref{fig:synth}.
The required value of $E_{\rm tot}$
for the source ``S'' is not largely changed even in this case.
A slightly softer index can be possible, but it should be significantly
harder than $\alpha=2$ as shown in Figure \ref{fig:single-pos}.

\begin{figure}[!h]
\centering
\epsscale{1.1}
\plotone{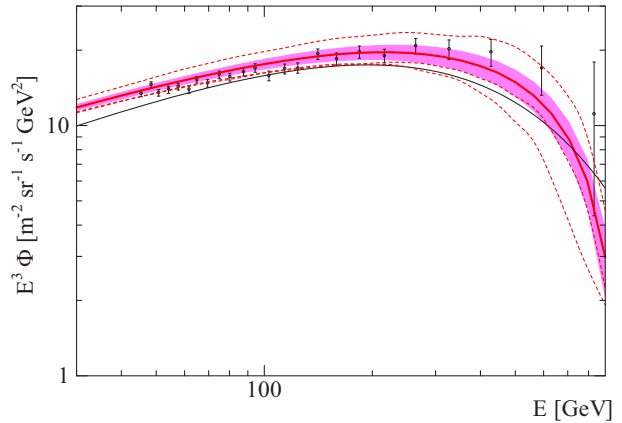}
\caption{Positron spectra obtained
from multiple pair release events with
a rate of $1.0 \times 10^{-2}~\mbox{kpc}^{-2}~\mbox{kyr}^{-1}$,
$K_0=3.0 \times 10^{28}~\mbox{cm}^2~\mbox{s}^{-1}$,
$q=1/3$, $E_{\rm tot}=6.5 \times 10^{47}$ erg,
$\alpha=1.9$, and $\gamma_{\rm max} m_{\rm e} c^2=10$ TeV.
We have assumed that no event within the last 500 kyr
contributes to the positron flux.
The red thick line is the average spectrum of 1000 trials.
The red dashed lines are three examples of the trials.
The black line is the model spectrum assumed in Figure \ref{fig:synth}.
\label{fig:pos}}
\end{figure}

Below we discuss the background pulsar component for the AMS-02 positron spectrum.
A model with multiple-source contributions is natural for the positron component, too.
We reproduce the model spectrum for this component
assumed in Figure \ref{fig:synth} with the same Monte Carlo method.
While several studies claimed that the radio pulsar birth rate is close to the SN rate
in our galaxy \citep{fau06},
not all SNe leave an active pulsar or a PWN
\citep[e.g. 1987A and Cassiopeia A, see][]{alp21,cha01}.
\citet{lor93} estimated the local pulsar birth rate density
as $0.6$--$1.2 \times 10^{-2}~\mbox{kpc}^{-2}~\mbox{kyr}^{-1}$,
which is $\sim 10$ times lower than the SN rate density we have assumed.
The fraction of high magnetic field or active pulsars
may be significantly lower \citep{vra04,lor06}.
Here we assume the birth rate of active pulsars as positron sources
as $1.0 \times 10^{-2}~\mbox{kpc}^{-2}~\mbox{kyr}^{-1}$.
As we have discussed in \S \ref{CRinj}, we consider that
positrons from the Geminga pulsar at $r=0.25$ kpc, whose age is 342 kyr,
do not largely contribute to the AMS-02 positron spectrum yet.
Thus, we assume that no event within the last 500 kyr
contributes to the positron spectrum,
which is preferable for the spectral peak at 200--300 GeV.

The results are shown in Figure \ref{fig:pos}.
The average spectral shape of the CRs from pulsars (red solid line)
is roughly consistent with the observed data from AMS-02
and the model spectrum other than the distribution of source ``S'' assumed
in Figure \ref{fig:synth} (the blue dashed line in Figure \ref{fig:synth}
and the black line in Figure \ref{fig:pos} are the same).
Also in this case, the spectral shape fluctuates depending on trials.
Even with a constant parameter set, we can find several trials
that produce a spectrum close to, or lower, than the model assumed
in Figure \ref{fig:synth}.
The required parameters are $\alpha=1.9$ and
$E_{\rm tot}=6.5 \times 10^{47}$ erg
(${\cal R} E_{\rm tot}=6.5 \times 10^{45}~\mbox{erg}~\mbox{kpc}^{-2}~\mbox{kyr}^{-1}$).
If we take into account the secondary positrons we have neglected in our calculations,
the injection index can be harder than 1.9.
The energy per pulsar is close to the Geminga's value
of $L_{\rm SD} t_{\rm age}$.
While we have assumed a constant $E_{\rm tot}$,
\citet{evo21} considered a Gaussian distribution in the initial spin period.
The average energy release in \citet{evo21} is
$1.5 \times 10^{47}$ erg with an average initial spin period of 100 ms.
This lower value than ours is due to the higher birth rate they adopted.
Thus, although the value of $E_{\rm tot}$ depends on the effective pulsar
birth rate, it may not largely exceed $10^{48}$ erg.

The injection spectral index $\alpha$ for the pulsar component is required
to be significantly harder than the index for SNRs.
This may be attributed to the intrinsic acceleration properties of pulsar CRs
or the escape process from PWNe.

The source ``S'' for the TeV sharp drop
requires a harder injection spectrum ($\alpha \geq 1.5$)
and a larger energy release ($E_{\rm tot}>10^{49}$ erg)
than those for the background pulsars.
Therefore, if this source really exists,
it belongs to another peculiar type of population
different from usual pulsars.
As discussed in \citet{koh16}, a peculiar type of event,
such as a supernova with an efficient secondary pair production
via interaction with molecular cloud, is required.

If the events like the source ``S'' occur with a rate
of $1/100$ of the SN rate,
the contribution from such sources to the positron spectrum at 300 GeV
is higher than the observed one.
To avoid the flux excess due to the stacking contribution of such
hard sources, the rate should be lower than $1/300$ of the SN rate.
This implies that the rate of source ``S''-type events is lower than once per
344 kyr within 2 kpc from us.
Our assumption of the event 420 kyr ago does not contradict this rate.
The upper limit of the rate is also roughly consistent with
the occurrence rate of broad-line Ibc SNe or hypernovae \citep{gue07}.

\subsection{Dispersion in the injection index}
\label{sub3}

While we have assumed a constant value for the injection index
of electron CRs from SNRs,
the real indices of the escaped electron CRs can have a dispersion.
From radio spectral indices in SNRs, \citet{cla76} found
a dispersion in the non-thermal electron index of $\sigma_\alpha \sim 0.3$.
The spectral index of electron CRs can be affected by the
escape process \citep{ohi10,ohi12} and/or radiative cooling
\citep{die19,mor21}.
The dispersion of the injection index can produce a modulation in the spectral shape.
The required property for the source ``S'' can be altered by such a modulation.

\begin{figure}[!h]
\centering
\epsscale{1.1}
\plotone{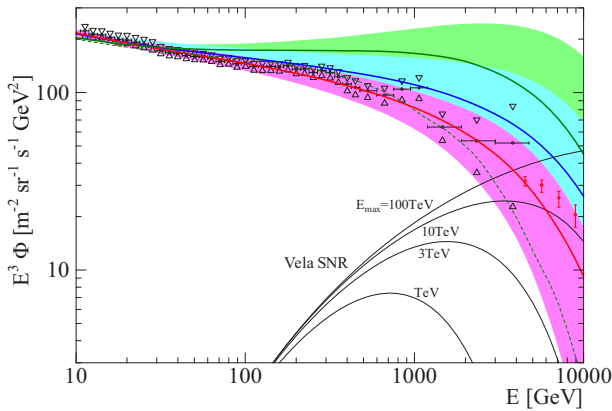}
\caption{The Monte Carlo simulation results
with Gaussian distribution for $\alpha$ ($\bar{\alpha}=2.58$).
Other parameters are the same as those in Figure \ref{fig:No-disp}:
SN rate of $7.7 \times 10^{-2}~\mbox{kpc}^{-2}~\mbox{kyr}^{-1}$,
$K_0=3.0 \times 10^{28}~\mbox{cm}^2~\mbox{s}^{-1}$,
$q=1/3$, $E_{\rm tot}=9.0 \times 10^{47}$ erg,
and $\gamma_{\rm max} m_{\rm e} c^2=10$ TeV.
The thick lines are the average spectra of 100 trials
with $\sigma_\alpha=0.15$ (red), 0.2 (blue), and 0.3 (green).
The spectral variation ($1\sigma$) is shown
with light green, cyan and magenta regions, for green, blue and red average spectra,
respectively.
The green dashed line is an example of the trials for $\sigma_\alpha=0.3$.
The red data points
are indirect measurements of electrons and positrons
by \citet{hess}.
The Vela SNR spectra (black) are calculated with the distance of 287 pc,
the age 10 kyr, $E_{\rm tot}=9.0 \times 10^{47}$ erg, $\alpha=2.52$,
and various $E_{\rm max}=\gamma_{\rm max} m_{\rm e} c^2$.
\label{fig:gaus}}
\end{figure}

Here we assume a Gaussian probability distribution for the index $\alpha$,
\begin{equation}
P(\alpha) \propto \exp{\left[ -(\alpha-\bar{\alpha})^2/(2 \sigma_\alpha^2)\right]},
\end{equation}
where constants $\bar{\alpha}$ and $\sigma_\alpha$ are the average index
and dispersion, respectively.
The number of trials to plot the average spectra
is 100 in Figure \ref{fig:gaus}.
This trial number is significant for our purpose.

As shown in Figure \ref{fig:gaus},
even with the dispersion in $\alpha$,
the spectral index around 10 GeV is almost the same
as in the case without dispersion.
However, a broader distribution (larger $\sigma_\alpha$)
tends to produce a harder spectrum above 100 GeV,
because contributions from harder components
dominate in the high energy region.
Sources with a softer index than $\bar{\alpha}$
do not contribute to the spectrum above $\sim$ TeV very much.
The average spectrum for $\sigma_\alpha=0.3$ seems too hard.
However, the dispersion in spectral shapes is large.
In the first 100 trials for $\sigma_\alpha=0.3$, we can find an example
consistent with the CALET spectrum, as shown by the green dashed line.

The data from the \citet{hess} plotted in Figure \ref{fig:gaus}
is unique data in this energy range.
However, the H.E.S.S. results are preliminary,
and show very small errors in spite the measurement being indirect.
They are qualitatively different data sets,
so we treat the H.E.S.S. data points as referential data in this paper.

\begin{figure}[!h]
\centering
\epsscale{1.1}
\plotone{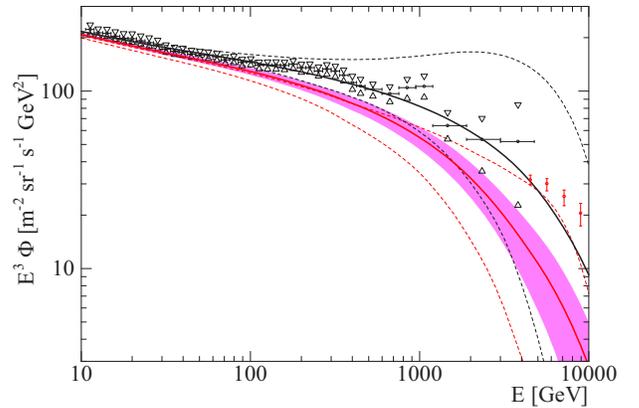}
\caption{The spectra with and without the contributions from nearby sources.
The thick black line is the same average spectrum
for $\sigma_\alpha=0.15$ in Figure \ref{fig:gaus}.
The black dashed lines are two examples of the trials.
The red lines show the results excluding events
within 500 pc and 100 kyr, for which
the average of 100 trilas (thick) and two examples of the trials (dashed) are shown.
The spectral variation ($1\sigma$) corresponding to the red solid line is shown
with the magenta region,
respectively.
\label{fig:gausexc}}
\end{figure}

Figure \ref{fig:gaus} also shows model spectra for the Vela SNR,
which has been discussed as a possibly dominant source in the TeV energy range
in previous studies \citep[e.g.,][]{kob04,kaw11,man19,for20}.
As we assumed in Figure \ref{fig:gaus}, if Vela's electron injection
is close to the average properties of other SNRs,
we may have already detected electron CRs from the Vela SNR,
depending on the cutoff energy.
However, the large dispersion in the spectrum makes it difficult
to definitely distinguish individual nearby sources.
Similarly to the leptonic CRs from pulsars,
a significant delay of CR escape may be possible even for the Vela SNR.
The delay can further diminish the flux of Vela.

\begin{figure}[!h]
\centering
\epsscale{1.1}
\plotone{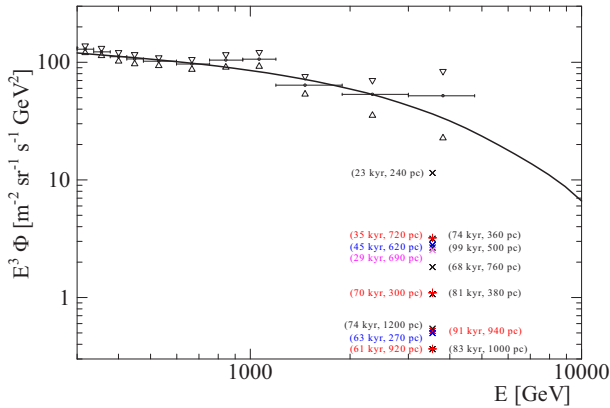}
\caption{An example of trials for $\sigma_\alpha=0.15$.
The parameters are the same as those in Figure \ref{fig:gaus}.
Contributions to the 3.5 TeV flux from individual sources
are denoted by crosses.
The top 14 flux contributions are plotted
with the age and distance of the sources.
To distinguish each contribution,
we change the color and symbol especially for overlapping points.
\label{fig:contri}}
\end{figure}

Figure \ref{fig:gausexc} shows the contribution of relatively distant sources
above $\sim$ TeV.
The red lines show cases where nearby sources
with ($r_i<500$ pc) $\cap$ ($t_i<100$ kyr) are artificially excluded.
The differences between red and black lines show the contribution
of the recent nearby sources.
Although the red lines tend to be lower than the fluxes shown with black lines,
Figure \ref{fig:gausexc} shows a significant contribution
of distant ($r_i>500$ pc) sources to the flux above $\sim$ TeV.
Such sources may be hard to identify observationally as a possible source.

Let us consider sources whose contribution to the last energy bin
at 3.5 TeV is larger than $3.5~\mbox{m}^{-2}~\mbox{sr}^{-1}~\mbox{s}^{-1}~
\mbox{GeV}^{2}$, which is 15\% of the lower value of the CALET fluxes with errors.
Hereafter, we call such sources ``primary contributors at 3.5 TeV''.
For the case of $\sigma_\alpha=0.15$,
the average source number of the primary contributors is 2.3.
The average age and the distance of the primary contributors at 3.5 TeV
are $39 \pm 22$ kyr and $400 \pm 150$ pc.
Therefore, a few distant ($\gtrsim 500$ pc) sources
largely contribute to the flux at 3.5 TeV.
However, only the contribution of such primary sources
does not account for most of the flux at 3.5 TeV in general.

Figure \ref{fig:contri} shows one typical example
of the flux contributions from individual sources.
In this case, the number of primary contributors at 3.5 TeV
is only one coming from a source at 240 pc.
This source contributes $\sim 1/3$ of the CALET flux at 3.5 TeV.
The residual $\sim 2/3$ of the CALET flux
are attributed to the sum of $\sim 10$ weak sources.

\subsection{Iroshnikov-Kraichnan Diffusion}
\label{IK}

\begin{figure}[!h]
\centering
\epsscale{1.1}
\plotone{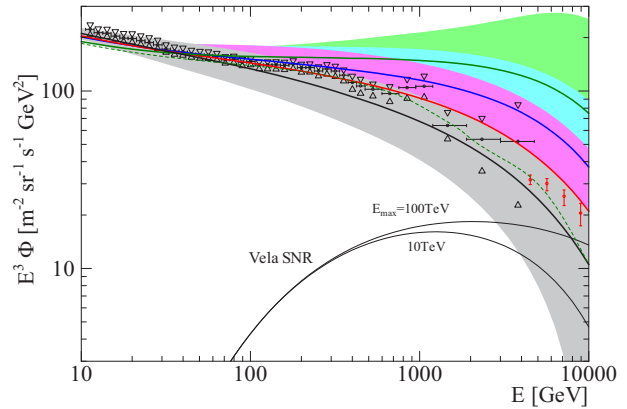}
\caption{Same as Figure \ref{fig:gaus} but for $q=1/2$.
The parameters are
SN rate of $7.0 \times 10^{-2}~\mbox{kpc}^{-2}~\mbox{kyr}^{-1}$,
$K_0=2.08 \times 10^{28}~\mbox{cm}^2~\mbox{s}^{-1}$,
$\bar{\alpha}=2.50$, $E_{\rm tot}=9.0 \times 10^{47}$ erg,
and $\gamma_{\rm max} m_{\rm e} c^2=10$ TeV.
The thick lines are the average spectra of 100 trials
with $\sigma_\alpha=0$ (black), 0.15 (red), 0.2 (blue), and 0.3 (green).
The spectral variation ($1\sigma$) is shown
with grey, light green, cyan and magenta regions, for green, blue, red,
and black average spectra,
respectively.
The green dashed line is an example of the trials for $\sigma_\alpha=0.3$.
The Vela SNR spectra (black) are calculated with the distance of 287 pc,
the age 10 kyr, $E_{\rm tot}=9.0 \times 10^{47}$ erg, $\alpha=2.52$,
and various $E_{\rm max}=\gamma_{\rm max} m_{\rm e} c^2$.
\label{fig:gaus2}}
\end{figure}

We have assumed Kolmogorov type diffusion.
In this section, to check the dependence on the spatial diffusion property,
we apply Iroshnikov--Kraichnan (IK) type
diffusion, $q=1/2$ with the diffusion coefficient
of $K_0=2.08 \times 10^{28}~\mbox{cm}^2~\mbox{s}^{-1}$
\citep{yua17,fan18b}.
The analytic estimate suggests a harder injection
spectral index $\Delta \alpha=(0.5-0.33)/2 \simeq 0.08$.
As expected, an index of $\bar{\alpha}=2.50$
agrees with the CALET spectrum as shown in Figure \ref{fig:gaus2}.
However, the implied proton injection index
($2.81 \pm 0.03-q=2.31 \pm 0.03$)
is significantly harder than $\bar{\alpha}=2.50$.
For the IK diffusion, the radiative cooling effect
on the escape process of electrons
\citep{die19,mor21} may have to be taken into account.

\begin{figure}[!h]
\centering
\epsscale{1.1}
\plotone{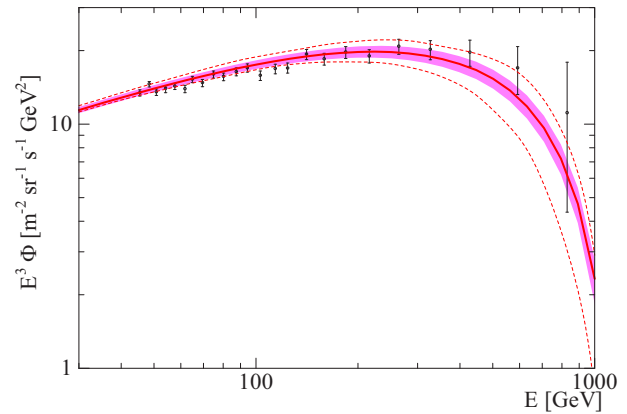}
\caption{Positron spectra for $q=1/2$ obtained
from multiple pair release events with
a rate of $1.0 \times 10^{-2}~\mbox{kpc}^{-2}~\mbox{kyr}^{-1}$,
$K_0=2.08 \times 10^{28}~\mbox{cm}^2~\mbox{s}^{-1}$,
$E_{\rm tot}=8.0 \times 10^{47}$ erg,
$\alpha=1.8$, and $\gamma_{\rm max} m_{\rm e} c^2=10$ TeV.
We have assumed that no event within the last 400 kyr
contributes to the positron flux.
The red thick line is the average spectrum of 1000 trials.
The spectral variation ($1\sigma$) is shown
with the magenta region.
The red dashed lines are two examples of the trials.
\label{fig:pos2}}
\end{figure}

The obtained spectra are similar to the cases for $q=1/3$
(Figure \ref{fig:gaus}).
To fit the CALET spectrum, we adopt a slightly lower birth rate
($7.0 \times 10^{-2}~\mbox{kpc}^{-2}~\mbox{kyr}^{-1}$,
${\cal R} E_{\rm tot}=6.3 \times 10^{46}~\mbox{erg}~\mbox{kpc}^{-2}~\mbox{kyr}^{-1}$)
than in the case for $q=1/3$.
A larger $\sigma_\alpha$ tends to produce a harder spectrum,
but we can find a few cases consistent with the CALET spectrum
even for $\sigma_\alpha=0.3$ from 100 trials
(see the green dashed line).
The contribution from Vela SNR for $q=1/2$
is lower than the model flux with $q=1/3$.

The positron spectrum is also reproduced by this diffusion model
with $\alpha=1.8$ as shown in Figure \ref{fig:pos2}.
A slightly larger energy release ($8.0 \times 10^{47}$ erg per pulsar,
${\cal R} E_{\rm tot}=8.0 \times 10^{45}~\mbox{erg}~\mbox{kpc}^{-2}~\mbox{kyr}^{-1}$)
is required compared to the Kolmogorov case.



\section{Summary}
\label{sec:sum}

Focusing on the electron and positron spectrum measured with CALET,
we have carried out Monte Carlo simulations of random births
of SNe and PWNe, and calculated the propagation of CRs
to the earth including the Klein--Nishina effect on the CR cooling.
This method can take into account the stochastic property
of births of nearby sources, which dominantly contribute to
the CR flux above TeV.
Our results are summarized as follows:

\begin{itemize}
\item A simple power-law energy dependence of the diffusion coefficient
can account for the CALET spectrum.
At least from the electron and positron spectrum alone,
we do not find strong evidence for the broken power-law dependence.
\item A cutoff at 10 TeV in the injection spectrum is favorable,
but the CALET spectrum does not necessarily require a cutoff.
\item In the case of Kolmogorov type diffusion, the local SN rate
is $7.7 \times 10^{-2}~\mbox{kpc}^{-2}~\mbox{kyr}^{-1}$
for an energy release per SN of $E_{\rm tot}=9.0 \times 10^{47}$ erg,
and a diffusion coefficient of $K_0=3.0 \times 10^{28}~\mbox{cm}^2~\mbox{s}^{-1}$.
In the case of Iroshnikov--Kraichnan type diffusion, the local SN rate
is $7.0 \times 10^{-2}~\mbox{kpc}^{-2}~\mbox{kyr}^{-1}$
for $E_{\rm tot}=9.0 \times 10^{47}$ erg and  $K_0=2.08 \times 10^{28}~\mbox{cm}^2~\mbox{s}^{-1}$.
\item In the case of Kolmogorov type diffusion, the average injection index is $2.58$ for SN electrons,
while the plausible index for SN protons is $2.48$.
On the other hand, in the case of Iroshnikov--Kraichnan type diffusion,
the indices are $2.50$, and $2.31$ for electrons and protons, respectively.
The slightly softer injection spectrum for electrons may be attributed to
the cooling effect at the escape from sources.
\item Depending on Monte Carlo trials, the spectrum above $\sim$ TeV
shows large variance.
A larger dispersion of the injection index tends to produce a harder
spectrum than the CALET spectrum especially above $\sim$ TeV.
A dispersion of the index larger than $\sigma_\alpha = 0.3$ is not favorable,
but we can find several spectra consistent with the CALET spectrum
from 100 Monte Carlo trials even for $\sigma_\alpha = 0.3$.
\item The CALET flux for 2--4 TeV does not contradict CRs from SNRs.
We do not need extraordinary sources to reproduce the flux in this energy range.
One to three nearby SNRs, possibly including the Vela SNR, can contribute to a few tens of percent
of the observed flux. However, it is not likely that a single source
accounts for almost all the flux in this energy range.
Many SNRs at a distance $\gtrsim 500$ pc within 100 kyr
account for the remaining several tens of percent of the observed flux.
\item The characteristic bump structure around 200--300 GeV in the CALET spectrum
is consistent with the additional component of electron--positron CRs
measured with AMS-02.
The energy injection rate of the pair CRs is $\gtrsim 5.0
\times 10^{45}~\mbox{erg}~\mbox{kpc}^{-2}~\mbox{kyr}^{-1}$.
The injection index is 1.8--1.9.
If the sources for positrons are PWNe, the delayed CR escape from
PWNe with a timescale of a few hundreds kyr is favorable.
The released energy per pulsar may not exceed $10^{48}$ erg.
\item The sharp drop at $\sim 1$ TeV in the CALET spectrum
can be reproduced by an extraordinary event with a CR release
$\sim 400$ kyr ago.
The required energy and the injection index are $>10^{49}$ erg
and 1.5, respectively.
Those parameters are apparently different from those required
for SNRs and PWNe.
The rate of such extraordinary events should be lower than $1/300$ of the rate of usual SNe.
\end{itemize}

The event rate of the sources required for the TeV sharp drop
is roughly consistent with the rate of broad-line Ibc SNe or hypernovae.
Alternatively, an accidentally efficient production of secondary electron and positron CRs
may have occurred in a SN event via interaction with molecular clouds \citep{koh16}.
The TeV sharp drop is still seen in the updated CALET spectrum \citep{epmCALET}.
We have tested our results with this preliminary data,
and confirmed that the results are unchanged.
A further inmprovement of the statistics in the CALET data will verify the significance of the sharp drop.

\begin{acknowledgments}
We gratefully acknowledge the support from the CALET collaboration team.
This work is supported by the joint research program
of the Institute for Cosmic Ray Research (ICRR),
the University of Tokyo.
\end{acknowledgments}


\begin{thebibliography}{}

\bibitem[Abeysekara et al.(2017)]{abe17}
Abeysekara, A. U., Albert, A., Alfaro, R., et al. 2017, Science, 358, 911
\bibitem[Abdollahi et al.(2017)]{abd17}
Abdollahi, S., Ackermann, M., Ajello, M., et al. 2017, \prd, 95, 082007
\bibitem[Ambrosi et al.(2017)]{dampe}
Ambrosi, G., An, Q., Asfandiyarov, R., et al. 2017, \nat, 552, 63
\bibitem[Adriani et al.(2011)]{PAMELA}
Adriani, O., Barbarino, G. C., Bazilevskaya, G. A., et al. 2011, \prl, 106, 201101
\bibitem[Adriani et al.(2017)]{cal17}
Adriani, O., Akaike, Y., Asano, K., et al. 2017, \prl, 119, 181101
\bibitem[Adriani et al.(2018)]{cal18}
Adriani, O., Akaike, Y., Asano, K., et al. 2018, \prl, 120, 261102
\bibitem[Adriani et al.(2019)]{cal19}
Adriani, O., Akaike, Y., Asano, K., et al. 2019, \prl, 122, 181102
\bibitem[Akaike \& Maestro(2021)]{B/CCALET}
Akaike, Y., \& Maestro, P. 2021, PoS(ICRC2021)112
\bibitem[Accardo et al.(2014)]{ams14pos}
Accardo, L., Aguilar, M., Aisa, D., et al. 2014, \prl, 113, 121101
\bibitem[Aguilar et al.(2014)]{ams14}
Aguilar, M., Aisa, D., Alpat, B., et al. 2014, \prl, 113, 221102
\bibitem[Aguilar et al.(2016)]{ams16}
Aguilar, M., Ali Cavasonza, L., Ambrosi, G., et al. 2016, \prl, 117, 231102
\bibitem[Aguilar et al.(2018)]{ams18}
Aguilar, M., Ali Cavasonza, L., Ambrosi, G., et al. 2018, \prl, 120, 021101
\bibitem[Aguilar et al.(2019)]{ams19}
Aguilar, M., Ali Cavasonza, L., Alpat, B., et al. 2019, \prl, 122, 101101
\bibitem[Alp et al.(2021)]{alp21}
Alp, D., Larsson, J., \& Fransson, C. 2021, \apj, 916, 76
\bibitem[Atoyan et al.(1995)]{ato95}
Atoyan, A. M., Aharonian, F. A., \& V\"olk, H. J. 1995, \prd, 52, 3265
\bibitem[Bertsch et al.(1992)]{ber92}
Bertsch, D. L., Brazier, K. T. S., Fichtel, C. E., et al. 1992, \nat, 357, 306
\bibitem[Blumenthal \& Gould(1970)]{blu70}
Blumenthal, G. R., \& Gould, R. J. 1970, Rev. Mod. Phys., 42, 237
\bibitem[Clark \& Caswell(1976)]{cla76}
Clark, D. H., \& Caswell, J. L. 1976, \mnras, 174, 267
\bibitem[Chang et al.(2008)]{ATIC}
Chang, J., Adams, J. H., Ahn, H. S., et al. 2008, \nat, 456, 362
\bibitem[Cholis \& Hooper(2014)]{cho14}
Cholis, I., \& Hooper, D., 2014, \prd, 89, 043013
\bibitem[Cholis et al. (2018)]{cho18}
Cholis, I., Karwal, T., \& Kamionkowski, M., 2018, \prd, 98, 063008
\bibitem[Delahaye et al.(2010)]{del10}
Delahaye, T., Lavalle, J., Lineros, R., Donato, F., \& Fornengo, N. 2010, \aap, 524, A51
\bibitem[Di Mauro et al.(2021)]{di21}
Di Mauro, M., Donato, F., \& Manconi, S. 2021, \prd, 104, 083012
\bibitem[Diesing \& Caprioli(2019)]{die19}
Diesing, R., \& Caprioli, D. 2019, \prl, 123, 071101
\bibitem[Ding et al. (2021)]{din21}
Ding, Y.-C., Nan, L., Wei, C.-C., Wu, Y.-L., \& Zhou, Y.-F., 2021, \prd, 103, 115010
\bibitem[DuVernois et al. (2001)]{HEAT}
DuVernois, M. A.., Barwick, S. W., Beatty, J. J., et al. 2001, \apj, 559, 296
\bibitem[Evoli et al.(2019)]{evo19}
Evoli, C., Aloisio, R., \& Blasi, P. 2019, \prd, 99, 103023
\bibitem[Evoli et al.(2020)]{evo20}
Evoli, C., Blasi, P., Amato, E., \& Aloisio, R. 2020, \prl, 125, 051101
\bibitem[Evoli et al.(2021)]{evo21}
Evoli, C., Amato, E., Blasi, P., \& Aloisio, R. 2021, \prd, 103, 083010
\bibitem[Fang et al.(2018)]{fan18}
Fang, K., Bi, X.-J., \& Yin, P.-F. 2018, \apj, 854, 57
\bibitem[Fang et al.(2018b)]{fan18b}
Fang, K., Bi, X.-J., Yin, P.-F., \& Yuan, Q. 2018, \apj, 863, 30
\bibitem[Fang et al.(2021)]{fan21}
Fang, K., Bi, X.-J., Lin, S.-J., \& Yuan, Q. 2021, Chin. Phys. Lett., 38, 039801
\bibitem[Faucher-Gigu\`ere \& Kaspi(2006)]{fau06}
Faucher-Gigu\`ere, C.-A., \& Kaspi, V. M. 2006, \apj, 643, 332
\bibitem[Chakrabarty et al.(2001)]{cha01}
Chakrabarty, D., Pivovaroff, M. J.,  Hernquist, L. E.,
Heyl, J. S., \& Narayan, R. 2001, \apj, 548, 800
\bibitem[Fornieri et al.(2020)]{for20}
Fornieri, O., Gaggero, D., \& Grasso, D. 2020, JCAP, 02(2020), 009
\bibitem[Fujita et al. (2009)]{fuj09}
Fujita, Y., Kohri, K., Yamazaki, R., \& Ioka, K. 2009, \prd, 80, 063003
\bibitem[Genolini et al.(2015)]{gen15}
Genolini, Y., Putze, A., Salati, P., \& Serpico, P. D. 2015, \aap, 580, A9
\bibitem[Guetta \& Della Valle(2007)]{gue07}
Guetta, D., \& Della Valle, M. 2007, \apj, 657, L73
\bibitem[Grimani(2007)]{gri07}
Grimani, C. 2007, \aap, 474, 339
\bibitem[H.E.S.S. Collaboration (2017)]{hess}
H.E.S.S. Collaboration 2017, 35th ICRC, Busan, South Korea, CRI215
\bibitem[Hooper et al.(2009)]{hoo09}
Hooper, D., Blasi, P., \& Serpico, P. D. 2009, JCAP, 01(2009), 025
\bibitem[Hooper et al.(2017)]{hoo17}
Hooper, D., Cholis, I., Linden, T., \& Fang, K. 2017, \prd, 96, 103013
\bibitem[Huang et al.(2018)]{hua18}
Huang, Z.-Q., Fang, K., Liu, R.-Y., \& Wang, X.-Y. 2018, \apj, 866, 143
\bibitem[Kappl et al.(2015)]{kap15}
Kappl, R., Reinert, A., \& Winkler, M. W. 2015, JCAP, 10(2015), 034
\bibitem[Kashiyama et al.(2011)]{kas11}
Kashiyama, K., Ioka, K., \& Kawanaka, N. 2011, \prd, 83, 023002
\bibitem[Kawanaka et al.(2010)]{kaw10}
Kawanaka, N., Ioka, K., \& Nojiri, M. M. 2010, \apj, 710, 958
\bibitem[Kawanaka et al.(2011)]{kaw11}
Kawanaka, N., Ioka, K., Ohira, Y., \& Kashiyama, K. 2011, \apj, 729, 93
\bibitem[Keane \& Kramer(2008)]{kea08}
Keane, E. F., \& Kramer, M. 2008, \mnras, 391, 2009
\bibitem[Kisaka \& Kawanaka(2012)]{kis12}
Kisaka, S., \& Kawanaka, N. 2011, \mnras, 421, 3543
\bibitem[Kobayashi et al.(2004)]{kob04}
Kobayashi, T., Komori, Y., Yoshida, K., \& Nishimura, J. 2004, \apj, 601, 340
\bibitem[Kohri et al.(2016)]{koh16}
Kohri, K., Ioka, K., Fujita, Y., \& Yamazaki, R. 2016, PTEP, 2016, 021E01
\bibitem[Lipari (2019)]{lip19}
Lipari, P. 2019, \prd, 99, 043005
\bibitem[Lorimer et al.(1993)]{lor93}
Lorimer, D. R., Bailes, M., Dewey, R. J., \& Harrison, P. A. 1993, \mnras, 263, 403
\bibitem[Lorimer et al.(2006)]{lor06}
Lorimer, D. R., Faulkner, A. J., Lyne, A. G., et al. 2006, \mnras, 372, 777
\bibitem[Malyshev et al.(2009)]{mal09}
Malyshev, D. Cholis, I., \& Gelfand, J. 2009, \prd, 80, 063005
\bibitem[Manconi et al.(2019)]{man19}
Manconi, S., Di Mauro, M., \& Donato, F. 2019, JCAP, 04(2019), 024
\bibitem[Mertsch(2018)]{mer18}
Mertsch, P. 2018, JCAP, 11(2018), 045
\bibitem[Morlino \& Celli(2021)]{mor21}
Morlino, G., \& Celli, S. 2021, arXiv:2106.06488
\bibitem[Ohira et al.(2010)]{ohi10}
Ohira, Y., Murase, K., \& Yamazaki, R. 2010, \aap, 513, A17
\bibitem[Ohira et al.(2012)]{ohi12}
Ohira, Y., Yamazaki, R., Kawanaka, N., \& Ioka, K. 2012, \mnras, 427, 91
\bibitem[Orusa et al.(2021)]{oru21}
Orusa, L., Manconi, S., Donato, F., \& Di Mauro, M. 2021, arXiv:2107.06300
\bibitem[Recchia et al. (2019)]{rec19}
Recchia, S., Gabici, S., Aharonian, F. A., \& Vink, J. 2019, \prd, 99, 103022
\bibitem[Shi \& Liu(2019)]{shi19}
Shi, Z.-D., \& Liu, S. 2019, \mnras, 485, 3869
\bibitem[Strong et al.(2007)]{str07}
Strong, A. W., Moskalenko, I. V., \& Ptuskin, V. S., 2007, Ann. Rev. Nuc. Part. Sci., 57, 285
\bibitem[Torii \& Akaike(2021)]{epmCALET}
Torii, S., \& Akaike, Y. 2021, PoS(ICRC2021)105
\bibitem[Torii et al.(2001)]{BETS}
Torii, S., Tamura, T., Tateyama, N., et al. 2001, \apj, 559, 973
\bibitem[Vranesevic et al.(2004)]{vra04}
Vranesevic, N., Manchester, R. N., Lorimer, D. R., et al. 2004, \apj, 617, L139
\bibitem[Yoshida et al. (2008)]{PPB}
Yoshida, K., Torii, S., Yamagami, T., et al. 2008, Adv. Space Res., 42, 1670
\bibitem[Yuan et al. (2017)]{yua17}
Yuan, Q., Lin, S.-J., Fang, K., \& Bi, X.-J. 2017, \prd, 95, 083007
\bibitem[Yuan et al. (2017)]{yus09}
Y\"uksel, H., Kistler, M. D., \& Stanev, T. 2009, \prl, 103, 051101
\bibitem[Zhang \& Cheng (2001)]{zha01}
Zhang, L., \& Cheng, K. S. 2001, \aap, 368, 1063
\end{thebibliography}
\end{document}